**Title**

# Non-Hermitian mode locking in an L-band monolithic InP dual-ring laser

**Authors**

Xiao Sun[1*], Mohanad Al Rubaiee[1], Jue Wang[1], John H. Marsh[1], Stephen J. Sweeney[1], Anthony E. Kelly[1], Lianping Hou[1*]

*Corresponding authors: Xiao.Sun@glasgow.ac.uk; Lianping.Hou@glasgow.ac.uk

**Affiliations**

1 James Watt School of Engineering, University of Glasgow, Glasgow, G12 8QQ, U.K

**Abstract**

Coupling between laser cavities provides a means of controlling optical modes through their gain, loss, and hybridization. Here we demonstrate electrically controlled mode locking in a monolithic InP dual-ring laser operating in the L-band without a dedicated saturable absorber. Independent biasing of the two rings accesses single-mode emission, mode locking, and hybridized-supermode operation, with a measured intermode beat frequency of 542 MHz in the latter regime. The laser generates 7.9 ps pulses at 18.11 GHz. A coupled-cavity model describes how differential nonlinear detuning modifies the modal gain–loss contrast, and time-domain simulations yield pulse durations and spectral widths close to those measured experimentally. In a device incorporating saturable absorbers, reverse bias enables 3.1 ps pulses and reduces the integrated timing jitter from 9.06 to 2.46 ps relative to the unbiased condition. These results connect electrical control of coupled-cavity states with picosecond pulse generation in a monolithic InP semiconductor platform for high-repetition-rate optical communications and microwave photonics.

## MAIN TEXT

## Introduction

Mode-locked semiconductor lasers combine electrical pumping, compact cavities, and high repetition rates, making them attractive sources for optical communications, microwave photonics, and time-resolved measurements[1-9]. In monolithic devices, pulse formation is commonly controlled through the interplay of saturable absorption, gain saturation, dispersion, and spectral filtering[10-13]. These processes determine how a short pulse can persist against amplification of the low-intensity background, while also setting its duration, chirp, and stability. A dedicated saturable absorber provides direct control over intensity-dependent loss[14,15], but coupled cavities offer another means of shaping the effective gain and loss experienced by the optical field. Understanding how this additional degree of freedom influences pulse formation could broaden the operating regimes accessible to integrated semiconductor lasers.

Non-Hermitian photonics provides a framework for controlling interacting optical modes through amplification, dissipation, and coherent coupling[16-18]. In coupled resonators, the gain–loss contrast and relative detuning determine both the complex eigenfrequencies and the distribution of the field between the cavities[19-21]. Parity–time (PT)-symmetric laser designs have used this dependence to favour selected modes, including demonstrations of single-mode emission in optically pumped microrings and electrically pumped integrated ring lasers[22-25]. Such results establish gain–loss engineering as a practical tool for spectral selection. Extending it to mode locking requires an additional connection: changes in the coupled-cavity states must produce a dynamical response that supports a pulse train, rather than merely selecting a stationary lasing frequency.

Several studies have begun to establish this connection. PT-symmetric mode locking was proposed through asymmetric coupling between longitudinal modes, demonstrating how non-Hermitian interactions can organize temporal emission[26]. A complementary theoretical approach uses the sensitivity of coupled microresonators near an exceptional point: a weak Kerr-induced refractive-index change can modify the effective cavity loss and provide an artificial saturable-absorption response[27]. More recently, experiments reported have demonstrated dissipative solitons in coupled fibre-ring laser cavities, with pulse stabilization attributed to selective PT-symmetry breaking by the Kerr nonlinearity[28]. On-chip coupled quantum-cascade ring lasers have enabled hybridized frequency combs with coexisting bright and dark solitons[29-31]. Together, these studies show that coupling can drive temporal self-organization, while the role of gain-medium dynamics depends on the physical platform.

In an electrically pumped quantum-well laser, the gain and refractive index respond to the same evolving carrier population[32,33]. Carrier depletion therefore couples amplitude and phase, while finite recovery introduces a dependence on the preceding optical intensity[34,35]. These effects coexist with Kerr nonlinearity and can modify the detuning, spectral envelope, and temporal profile of a pulse[36]. Moreover, changing the current in either ring affects both amplification and resonance frequency. Translating coupled-cavity pulse formation to an InP platform consequently requires examining how gain–loss contrast, detuning, and nonlinear carrier dynamics act together. A central question is whether independent electrical control of the rings can access a mode-locked regime without a dedicated saturable absorber, and how the resulting pulses relate to the coupled-cavity mode structure.

Here we demonstrate electrically controlled mode locking in a monolithic InP dual-ring laser operating in the L-band. Independent biasing of the rings accesses predominantly single-mode emission, a finite mode-locking window, and hybridized-supermode operation. Without a dedicated saturable absorber, the device generates 7.9 ps pulses at 18.11 GHz. We interpret these observations using a non-Hermitian coupled-cavity model in which differential nonlinear detuning modifies the relative modal growth rates. Time-domain simulations give pulse durations and spectral widths close to those measured, and fits incorporating carrier-induced chirp describe the measured spectral and autocorrelation envelopes. Integrated saturable absorbers provide further pulse control, enabling 3.1 ps pulses and reduced timing noise. These results establish coupled-cavity gain–loss engineering as a route to electrically controlled picosecond pulse generation without a dedicated saturable absorber, offering a compact AlGaInAs/InP semiconductor platform for high-repetition-rate optical communications and microwave photonics.

## Result

### *Coupled-cavity model and nonlinear response*

The device comprises two evanescently coupled InP ring lasers with separate injection contacts and an integrated semiconductor optical amplifier (SOA) for output amplification (Fig. 1a). Both rings use the same AlGaInAs/InP multiple-quantum-well (MQW) waveguide structure, with a ridge width of 2 μm and an inter-ring gap of 1.7 μm (Fig. 1b). Varying the two injection currents changes the gain–loss contrast, resonance detuning, and intracavity power distribution. We refer to the more strongly pumped cavity as the gain ring and the second cavity as the loss ring; these labels specify their roles rather than fixed values of their net gain.

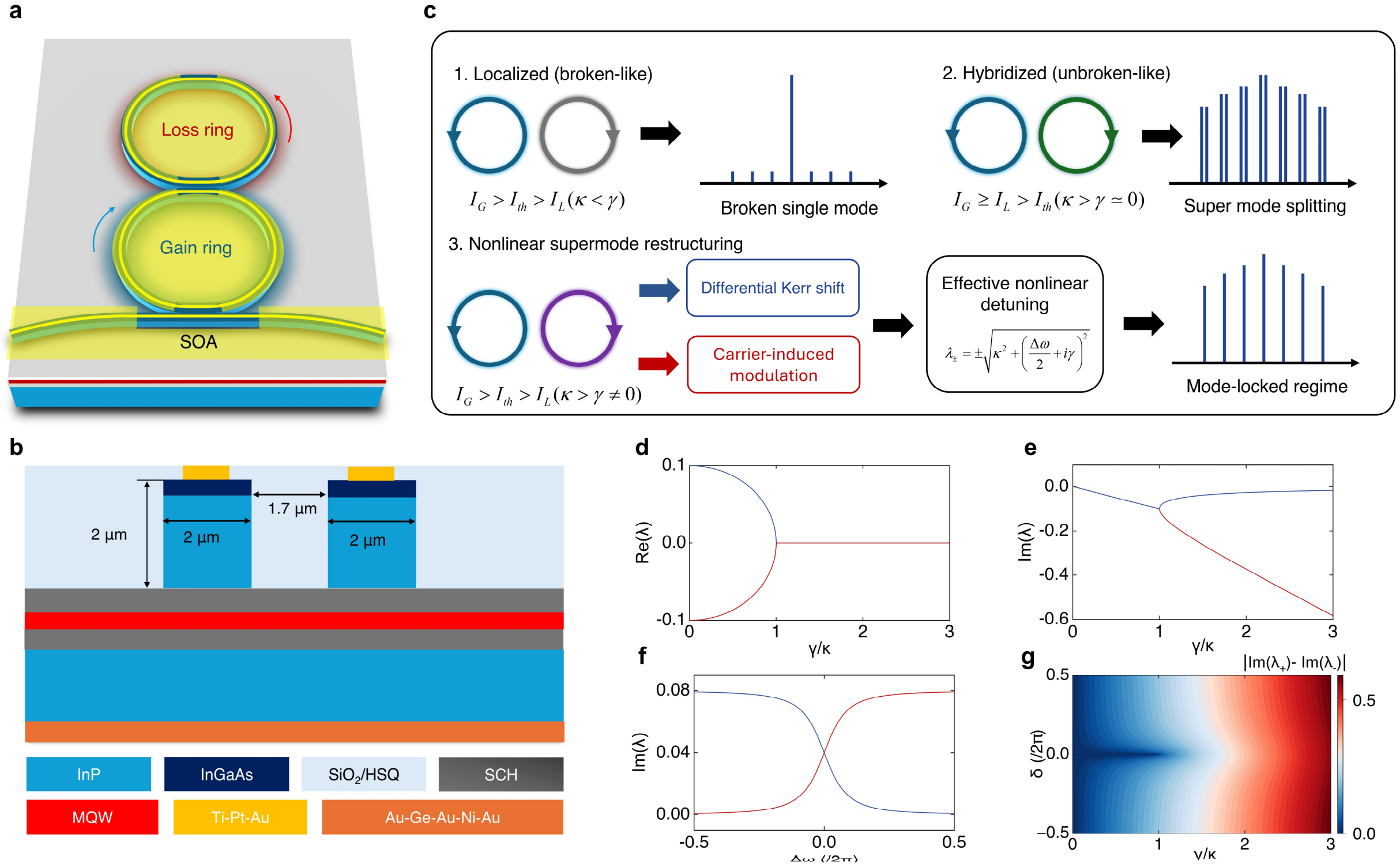

**Fig. 1: Device architecture and principle of nonlinear non-Hermitian mode locking. a**, Monolithic InP dual-ring laser with independently biased rings and an output SOA. **b**, Cross-section of the coupled AlGaInAs/InP ridge waveguides, showing the epitaxial layers, contacts, ridge dimensions, and coupling gap. **c**, Schematic of localized, hybridized, and nonlinear coupled-cavity states. Kerr and carrier-induced phase shifts contribute to differential nonlinear effect. The diagram illustrates the proposed interpretation of the mode-locked regime. **d, e**, Calculated real (d) and imaginary (e) parts of the linear supermode eigenvalues versus normalized gain–loss contrast $\gamma/\kappa$, with the exceptional point at $\gamma/\kappa = 1$. The nominal inter-ring power coupling is $K = 0.01$. Blue and red curves show the two eigenvalue branches. **f**, Imaginary parts of the supermode eigenvalues versus differential detuning at finite gain–loss contrast. **g**, Magnitude of the imaginary eigenvalue splitting versus gain–loss contrast and differential detuning. SOA, semiconductor optical amplifier; MQW, multiple quantum well; SCH, separate-confinement heterostructure; HSQ, hydrogen silsesquioxane.

For a pair of resonances with zero static detuning, the linear coupled-cavity eigenfrequencies are

$$\lambda_{\pm} = \bar{\omega} + i\bar{\sigma} \pm \sqrt{\kappa^2 + \left(\frac{\Delta\omega}{2} + i\gamma\right)^2} \tag{1}$$

Here, $\bar{\omega}$ is the common resonance frequency, κ is the coherent coupling rate, and $\bar{\sigma} = (\sigma_G + \sigma_L)/2$ and $\gamma = (\sigma_G - \sigma_L)/2$ represent the mean net field-growth rate and half the difference between the signed net field-growth rates of the gain and loss rings, respectively. The total relative detuning is $\Delta\omega$. Exact PT symmetry requires $\bar{\sigma} = 0$; after removal of the common term, the model has an unbroken-like regime for $|\gamma| < \kappa$ and a broken-like regime for $|\gamma| > \kappa$, separated by an exceptional point at $|\gamma| = \kappa$ (Fig. 1c–e). These conditions refer to the zero-detuning model.

At finite $\gamma$, a real differential detuning $\Delta\omega$ can split the imaginary parts of the eigenvalues of Eq. (1), changing the relative modal growth rates (Fig. 1f, g). An intensity-dependent detuning can therefore provide nonlinear gain–loss discrimination between the supermodes. Near the exceptional point, this response can be sensitive to weak detuning, although the full eigenvalue expression is required when a perturbative expansion is no longer valid (Supplementary Note 2). This two-mode analysis identifies a possible contribution to pulse formation; longitudinal-mode dynamics, gain saturation, and carrier response are included in the time-domain model to examine the resulting pulsed states. Under weak inter-ring coupling, each lasing ring can be treated approximately as an independently gain-clamped cavity. When both rings sustain lasing ($I_G > I_L > I_{th}$), gain clamping[37] brings the modal gain in each ring cavity close to loss, which makes $\gamma \rightarrow 0$, $\Delta\omega$ detuning changes only the real eigenvalue separation in this model, the output modes is hybridized (Fig. 1c).

Dissipative soliton pulses in passively mode-locked lasers are sustained by a dual balance: between dispersion and nonlinearity, and between gain and loss[38]. In this dual-ring regime, the mode-locking measurements use a gain-ring current above the standalone threshold and a loss-ring current below it ($I_G > I_{th} > I_L$). This bias configuration allows a finite gain–loss contrast and an intracavity power imbalance. The model includes a differential nonlinear detuning with Kerr and carrier-induced contributions. Mode locking in PT-symmetric dispersive resonators relies on Kerr-induced detuning that can selectively break PT symmetry[27,39]. An imbalance between the intracavity powers of gain-ring power $P_G$ and loss-ring power $P_L$ introduces a differential detuning, $\Delta\omega_K \propto (P_G - P_L)$.This mechanism reduces the effective loss experienced by strong pulses in the gain ring coupled to weaker pulses in the loss ring. The carrier-induced detuning $\Delta\omega_H \propto \int_{t_0}^{t} P(t')dt'$ differs from instantaneous intensity-dependent transmission. The interplay of Kerr nonlinearity, dispersion, and non-Hermitian coupling can provide intensity-dependent pulse shaping and gain–loss discrimination analogous to the action of a saturable absorber. Carrier depletion further introduces gain and phase modulation with memory, which can cause the pulse profile to deviate from an ideal $\mathrm{sech}^2$ shape (Supplementary Note 2).

*Device fabrication and measurement*

The fabricated device contains two rings of 700 μm radius connected by a nominal 1% power coupler (Fig. 2a). The gain ring is coupled to a 2300 μm long SOA waveguide through a nominal 10% power coupler. The coupling sections between the rings and the SOA are 160 μm long, with an inter-ring waveguide gap of 1.7 μm and a 1 μm gap between the gain ring and the SOA waveguide. Scanning electron microscopy (SEM) shows parallel ridges across the coupling section (Fig. 2b, c).

Cross-sectional SEM images show the two ridges and their separate contacts (Fig. 2d, e). Each ridge carries a 1 μm-wide microelectrode, allowing the rings to be biased independently despite their close optical proximity. Fabrication details are given in Methods and Supplementary Note 4. The calculated intensity distributions of the two transverse coupler modes are shown in Fig. 2f, g, with horizontal and vertical profiles for mode 1 in Fig. 2h, i. The calculated effective mode area of a single ridge is approximately 3.8 $\mu m^2$.

The device was mounted on a thermoelectrically controlled (TEC) stage, with separate current sources for the two rings and the SOA (Fig. 3a). At $I_L$ = 75 mA, varying $I_G$ produces stepwise changes in the optical spectrum, consistent with current-dependent resonance shifts and inter-ring detuning (Fig. 3b). A complementary scan of $I_L$ at $I_G$ = 450 mA reveals a finite mode-locking window between 72 and 90 mA (Fig. 3c). At $I_L$ = 0 mA, the emission is predominantly broken-single mode (Fig. 3d). At $I_L$ = 75 mA, the optical spectrum broadens (Fig. 3e), and the radiofrequency (RF) spectrum shows a fundamental beat note at 18.11 GHz and a second harmonic near 36.20 GHz (Fig. 3h). A Lorentzian fit to the fundamental peak gives a full-width at half-maximum (FWHM) of 804 kHz (Fig. 3i). These spectral features, together with the temporal measurements below, identify the mode-locked state.

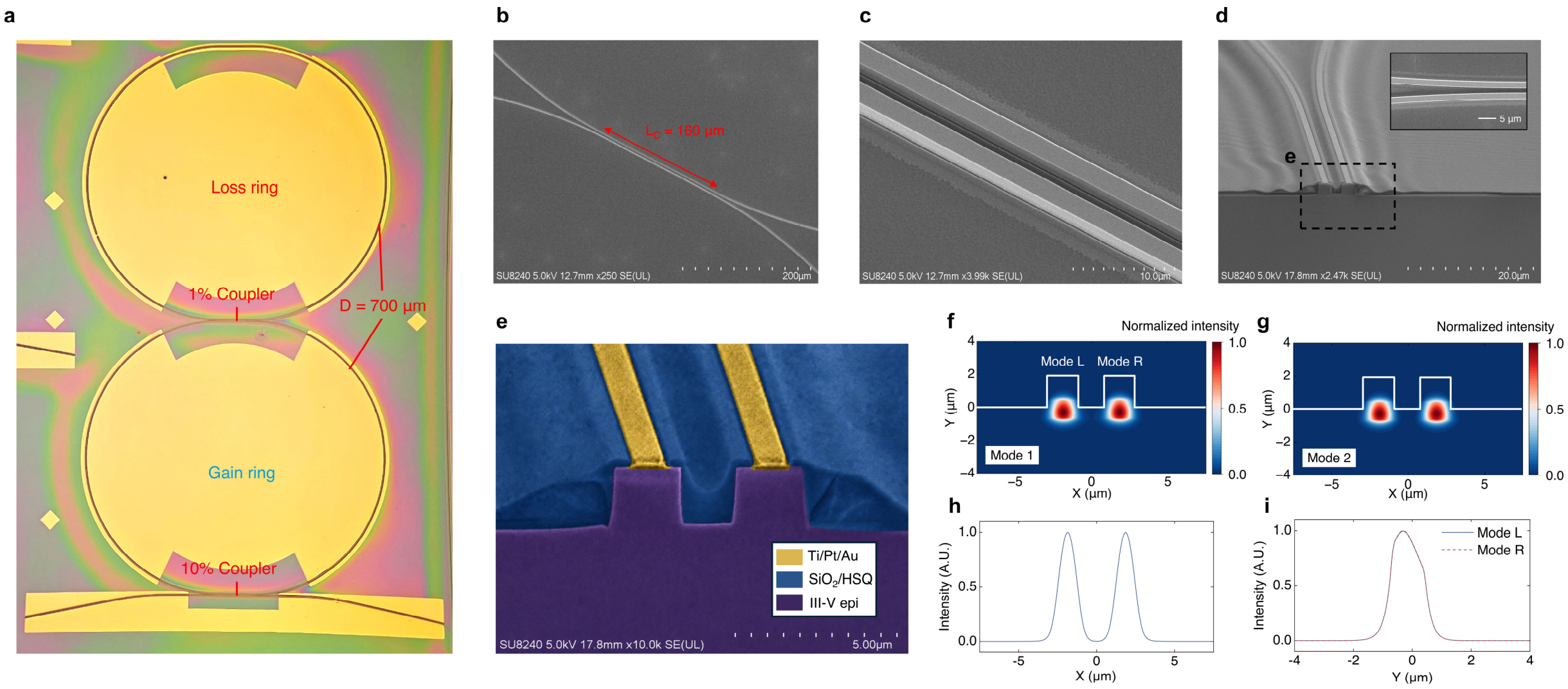

**Fig. 2: Device geometry, fabricated coupling structures and optical-mode confinement. a,** Optical micrograph of the dual-ring laser, showing the 700 μm diameter rings, nominal 1% inter-ring power coupler, and nominal 10% output coupler to the SOA waveguide. **b**, SEM image of the inter-ring coupling section, with an interaction length of approximately 160 μm. **c**, Magnified view of the parallel ridges in the directional coupler. **d**, Cross-sectional SEM image of the coupling region; inset, plan view of the parallel ridges. **e**, False-colour detail of the region outlined in **d**, highlighting the contacts, passivation, and semiconductor ridges. **f, g,** Calculated transverse intensity distributions of two coupler eigenmodes 1 and 2. **h, i**, Normalized horizontal (**h**) and vertical (**i**) intensity profiles of mode 1.

At $I_L$ = 108 mA, both rings are in a lasing state; the optical and RF spectra indicate a different coupled-cavity state (Fig. 3f, j). In addition to the beat note near the cavity free spectral range (FSR), a low-frequency peak appears at 542 MHz (Fig. 3k), consistent with beating between hybridized supermodes. Under the assumptions of negligible static detuning and a gain–loss contrast small compared with $\kappa$, this splitting gives $\kappa/(2\pi) \approx 271$ MHz, close to the value expected for the nominal 1% power coupler. The conversion from beat frequency to coupling rate is conditional on these assumptions because both detuning and gain–loss contrast affect the eigenfrequency separation.

The I-P curves show a threshold near $I_{th}$ = 100 mA and increasing output above threshold for the SOA currents examined (Fig. 3g). To place the mode-locking window relative to the model transition, we use the approximate loss and gain–current calibration described in Supplementary Note 1. This analysis adopts an internal waveguide loss of $3 \pm 0.5$ cm$^{-1}$ and uses below-threshold Hakki–Paoli net-modal-gain measurements[40]. With an approximately linear gain–current relation, it places the zero-detuning exceptional-point condition near $I_L$ = 67 mA. The measured window at 72–90 mA lies on the higher-current side of this estimate. This comparison is consistent with operation near a non-Hermitian transition, while the calibration and residual inter-ring detuning limit the precision of the exceptional-point assignment.

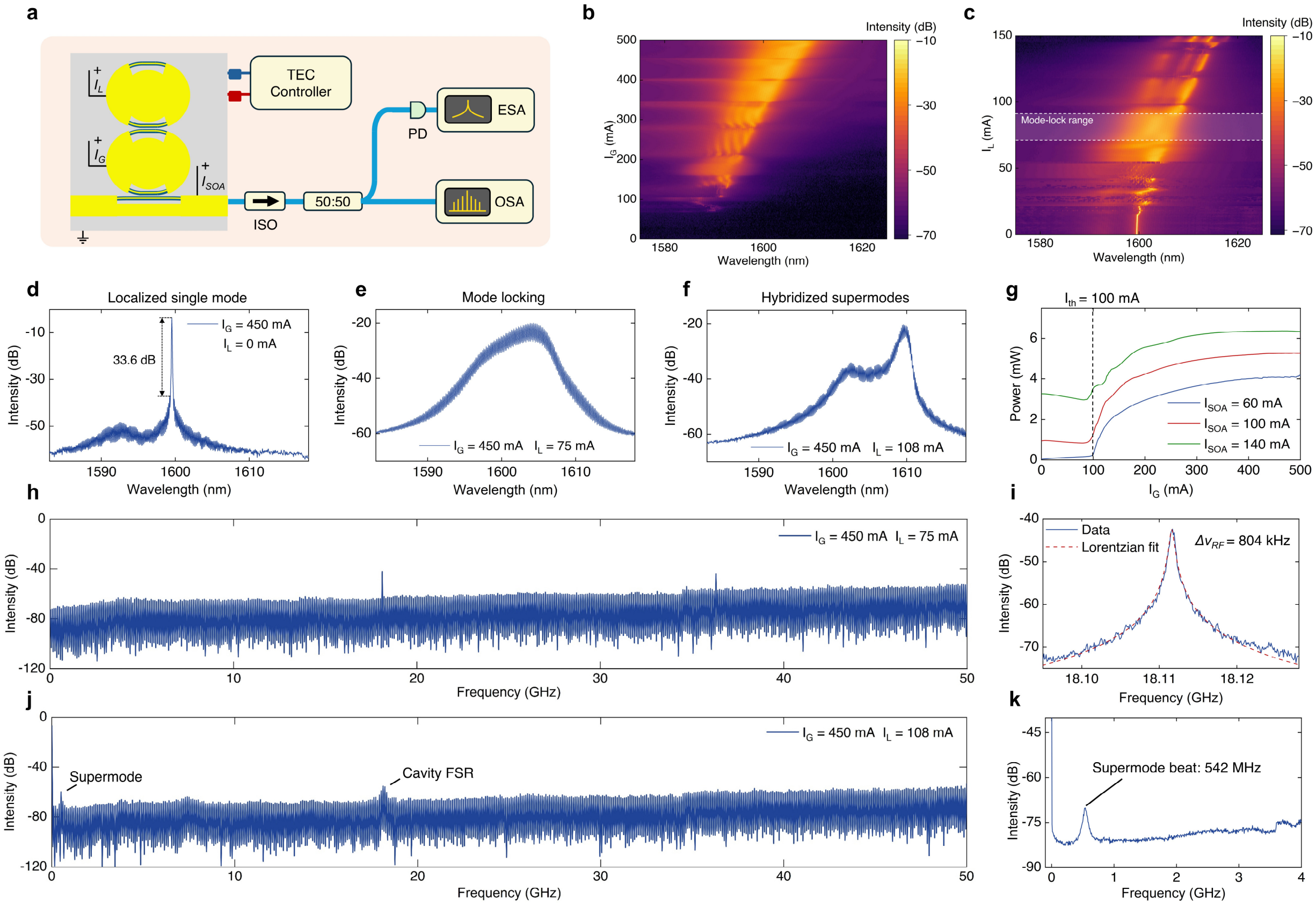

**Fig. 3: Electrical control of the optical and RF states of the coupled-ring laser. a,** Setup for optical and RF characterization, with independent biasing of the rings and SOA. **b**, Optical spectrum versus gain-ring current $I_G$ at $I_L$ = 75 mA. **c**, Optical spectrum versus loss-ring current $I_L$ at $I_G$ = 450 mA. Dashed lines mark the reported mode-locking window, $I_L$ = 72–90 mA. **d–f**, Optical spectra at $I_G$ = 450 mA and $I_L$ = 0 mA (**d**), 75 mA (**e**) and 108 mA (**f**), showing predominantly broken-single-mode, mode-locked and hybridized-supermode operation, respectively. **g**, Output power versus $I_G$ at SOA currents of 60, 100 and 140 mA. h, RF spectrum at $I_G$ = 450 mA and $I_L$ = 75 mA, with the fundamental at 18.11 GHz and the second harmonic near 36.20 GHz. **i**, Fundamental RF peak and Lorentzian fit, giving a FWHM of 804 kHz. **j**, RF spectrum at $I_G$ = 450 mA and $I_L$ = 108 mA, showing the beat notes near the cavity FSR and at low frequency. **k**, Enlarged low-frequency trace, with a peak at 542 MHz attributed to supermode beating. TEC, thermoelectric controller; ISO, optical isolator; OSA, optical spectrum analyser; PD, photodetector; ESA, electrical spectrum analyser

At $I_G$ = 450 mA and $I_L$ = 75 mA, the optical spectrum is centred near 1,605 nm and has a FWHM of 6.9 nm (Fig. 4a). The intensity-autocorrelation trace shows a pulse spacing of approximately 55 ps, consistent with the 18.11 GHz RF beat note (Fig. 4b, c). The central autocorrelation peak has a FWHM of 12.2 ps, with an estimated pulse duration of 7.9 ps (Fig. 4d). Fits incorporating the carrier-induced chirp correction described in Supplementary Note 2 closely reproduce both the optical spectral envelope and the intensity-autocorrelation profile (Fig. 4a, d). The fitted envelopes therefore account for the carrier-induced modification of the pulse, rather than assuming an ideal unchirped $sech^2$ profile. The resulting time–bandwidth product (TBP) of approximately 6.3 is consistent with strongly chirped emission. The measured pulse duration (7.9 ps) and spectral bandwidth (6.9 nm) are in close agreement with the values of 7.8 ps and 7.07 nm obtained from the time-domain simulations incorporating carrier dynamics (Supplementary Note 2 and Supplementary Fig. 2e, f). This agreement in both the pulse characteristics and envelope shape supports the validity of the coupled-cavity model, including the effects of carrier-induced phase modulation.

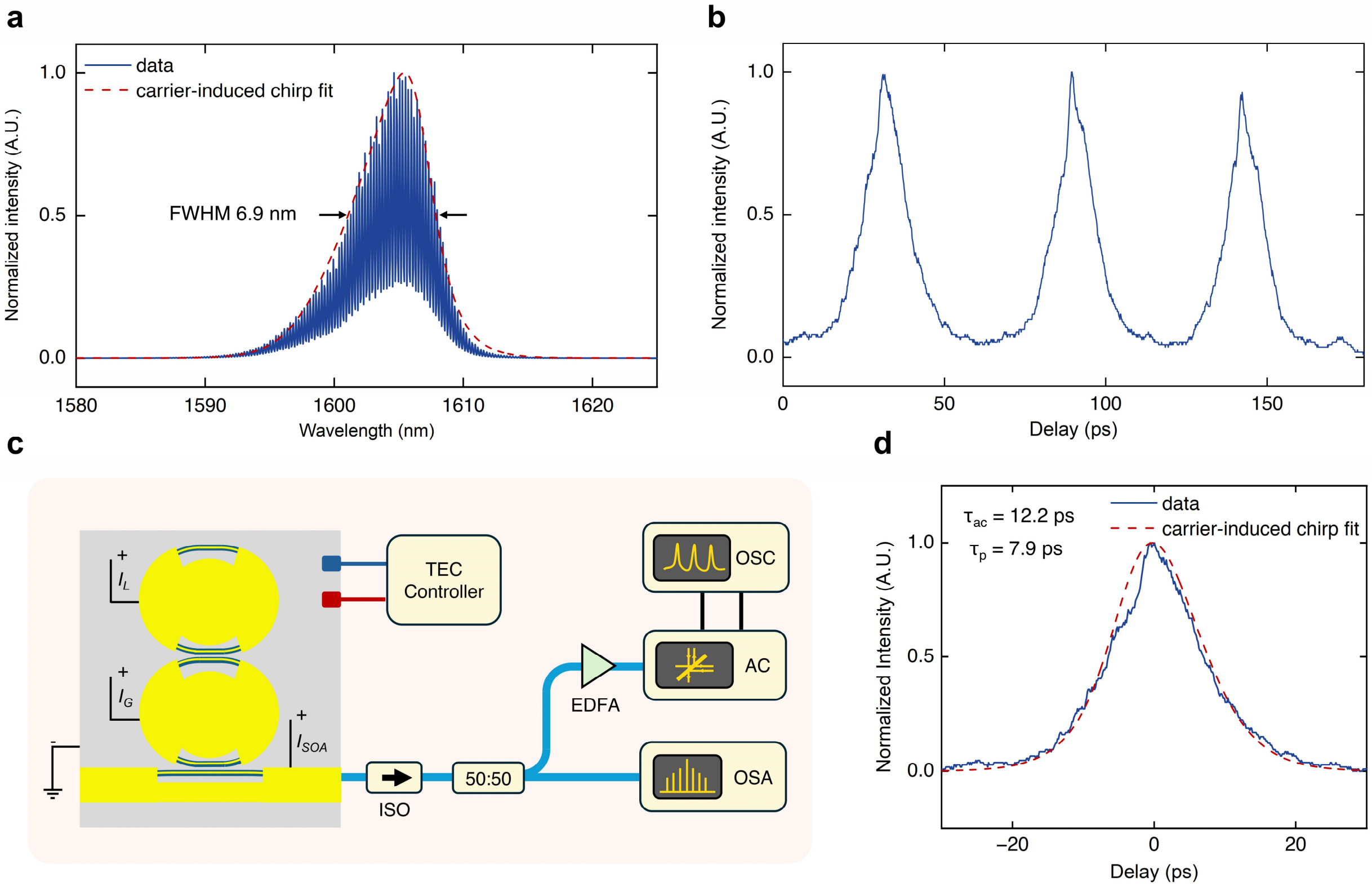

**Fig. 4: Temporal characterization of the electrically controlled mode-locked state. a,** Optical spectrum at $I_G$ = 450 mA and $I_L$ = 75 mA, with a FWHM of 6.9 nm, and an envelope fit incorporating the carrier-induced chirp correction described in Supplementary Note 2. **b**, Intensity autocorrelation over a wide delay range, showing a pulse spacing of approximately 55 ps. **c**, Temporal-characterization setup. The signal sent to the autocorrelator passes through an erbium-doped fibre amplifier (EDFA), whereas the optical spectrum is measured in the other branch. **d**, Central autocorrelation peak and fit incorporating the same carrier-induced chirp correction. The autocorrelation FWHM is 12.2 ps, with an estimated pulse duration of 7.9 ps. The fitted profiles in (a) and (d) closely reproduce the corresponding experimental measurements. OSC, oscilloscope; AC, autocorrelator.

### *Pulse optimization with integrated saturable absorbers*

We next examined a device containing a 35 μm long, electrically isolated saturable-absorber (SA) section in each ring (Fig. 5a). The SA introduces an additional mode-locking mechanism by preferentially transmitting intense pulses over the weak background, allowing saturable absorption to act alongside the nonlinear coupled-cavity dynamics in sustaining mode locking. The two SA biases provide additional control through bias-dependent absorption. With the gain-ring SA at $V_G = -1.5$ V and the loss-ring SA at $V_L = -1.1$ V, the optical bandwidth reaches 10.1 nm (Fig. 5b), and the RF spectrum shows a fundamental peak near 18.15 GHz with a FWHM of 119 kHz (Fig. 5c). The autocorrelation peaks are separated by 55.1 ps (Fig. 5d). The central peak has a FWHM of 4.8 ps, with an estimated pulse duration of 3.1 ps (Fig. 5e). Applying the carrier-induced chirp correction described in Supplementary Note 2 also gives close fits to the optical spectral envelope and autocorrelation profile of the SA-integrated device (Fig. 5b, e).

Reverse bias also reduces the measured single-sideband noise (Fig. 5f). The reported root-mean-square timing jitter, integrated from 10 kHz to 100 MHz, decreases from 9.06 ps with both SA sections at 0 V to 2.46 ps under reverse bias. Thus, absorber bias provides a further means of controlling pulse duration and timing noise in the coupled-ring device.

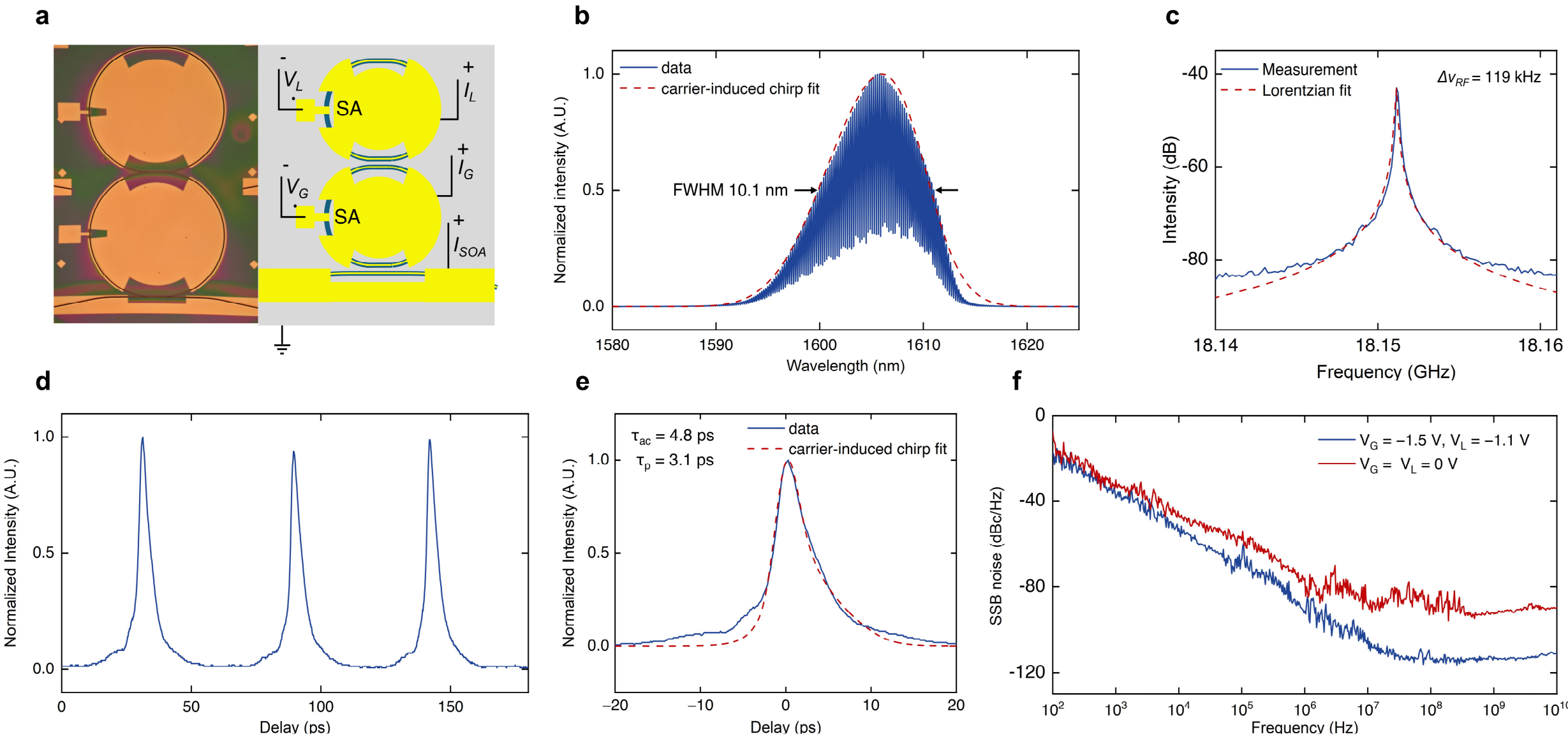

**Fig. 5: Saturable-absorber-assisted optimization of the nonlinear mode-locked state. a,** Optical micrograph and schematic of the dual-ring device with an independently biased, 35 μm long SA section in each ring. **b–e**, Measurements with the gain-ring and loss-ring SAs at $V_G = -1.5$ V and $V_L = -1.1$ V, respectively. b, Optical spectrum with a FWHM of 10.1 nm and an envelope fit incorporating the carrier-induced chirp correction described in Supplementary Note 2. **c**, Fundamental RF peak near 18.15 GHz and Lorentzian fit, giving a FWHM of 119 kHz. **d**, Wide-delay intensity autocorrelation with a peak spacing of 55.1 ps. **e**, Central autocorrelation peak and fit incorporating the same carrier-induced chirp correction. The autocorrelation FWHM is 4.8 ps, with an estimated pulse duration of 3.1 ps. The corrected fits closely reproduce the spectral and autocorrelation profiles in b and e. **f**, Single-sideband noise spectra with both SAs at 0 V and with the reverse biases used in b–e. The corresponding timing jitter integrated from 10 kHz to 100 MHz is 9.06 and 2.46 ps, respectively.

## Discussion

This work shows that electrical control of two coupled semiconductor rings accesses a finite mode-locking window between predominantly single-mode and hybridized-supermode operation. Additional current scans show that this window persists across gain-ring biases from 410 to 500 mA and shifts with bias (Supplementary Note 3). The coupled-cavity model offers a physical interpretation: differential nonlinear detuning changes the relative supermode growth rates when a finite gain–loss contrast is present. The proximity of the measured operating window to the estimated zero-detuning transition, together with the simulated pulse characteristics, supports this interpretation. Coupled semiconductor ring lasers have recently been shown to support hybridized bright–dark soliton frequency combs, while passive mode locking based directly on PT-symmetric coupled cavities has been demonstrated through Kerr-induced selective symmetry breaking.

Pulse generation without a dedicated SA and the shorter pulses obtained in the SA-integrated device show that coupled-cavity operation can be combined with absorber-based pulse control. The substantial time–bandwidth product leaves scope for pulse compression, provided that the output phase can be characterized and compensated. More broadly, independent electrical control of coupled resonators provides a practical way to explore the relation between non-Hermitian mode structure and ultrafast dynamics in an integrated semiconductor laser. Microresonator soliton frequency combs provide multiwavelength light sources for massively parallel coherent optical communications[41,42]. Our result offers a compact architecture for electrically ultrafast sources, with potential applications in optical communications and microwave photonics.

## Materials & Methods

### Device fabrication

The fabrication sequence is summarized in Supplementary Note 4. The epitaxial wafer was grown on an InP substrate by metal–organic vapour-phase epitaxy (MOVPE). Ring waveguides were patterned by electron-beam lithography using an EBPG5200 system. Negative-tone hydrogen silsesquioxane (HSQ) served as the resist and hard mask for inductively coupled plasma (ICP) etching with a $Cl_2/CH_4/H_2$ gas mixture in an Oxford PlasmaPro 300 system. Subsequent processing comprised $SiO_2$ deposition by plasma-enhanced chemical vapour deposition (Oxford PlasmaPro 100), HSQ passivation, contact-window opening, p-contact deposition, substrate thinning and n-contact deposition. SEM images were acquired with a Hitachi SU8240 at an accelerating voltage of 5 kV.

### Measurements setup

Optical spectra were acquired with an optical spectrum analyser (Agilent 86140B) at a resolution of 0.06 nm. Output light power was measured using a power meter (PM100D). The RF signal was detected using a photodetector (CXPDV2320R) and analysed with an electrical spectrum analyser (Keysight XA3-class). The resolution bandwidth (RBW) is 10 kHz. Single-sideband noise spectra $S_\Phi$(f) were measured using an electrical spectrum analyser (Keysight XA3-class) at a carrier frequency of 18 GHz. Timing jitter was integrated over offset frequencies from 10 kHz to 100 MHz. The conversion from the measured noise spectrum to root-mean-square timing jitter was

$$\sigma_t = \frac{1}{2\pi FSR}\sqrt{\int_{f_1}^{f_2} S_\phi(f)df} \quad (2)$$

Intensity autocorrelations were recorded using an autocorrelator (FR-103XL). An EDFA (AEDFA-L-EX2-DWDM-27-R) was placed in the autocorrelation branch, as shown in Fig. 4c.

## Acknowledgments

We would like to acknowledge the staff of the James Watt Nanofabrication Centre at the University of Glasgow for their help in fabricating the devices.

## Funding

Innovation Funding Service, UK Government (Innovate UK) (10056367)
Innovation Funding Service, UK Government (Innovate UK) (10158237)

## Competing interests

All authors declare they have no competing interests

## Data and materials availability

All data are available in the main text or the supplementary materials

# Supplementary Materials for

## Non-Hermitian mode locking in an L-band monolithic InP dual-ring laser

Xiao Sun[1], *et.al.*
[1] James Watt School of Engineering, University of Glasgow, Glasgow, G12 8QQ, U.K
Xiao.Sun@glasgow.ac.uk; Lianping.Hou@glasgow.ac.uk

**This PDF file includes:**

## Supplementary Note 1: Characterization of the modal loss

We characterized the modal gain and loss near the lasing threshold to estimate the gain–loss contrast in the coupled-ring model. Supplementary Fig. S1a shows spectra recorded around the threshold current ($I_{\mathrm{th}}$ = 100 mA), with the loss-ring current fixed at $I_{\mathrm{L}}$ = 75 mA. The output-waveguide semiconductor optical amplifier (SOA) was biased at $I_{\mathrm{SOA}}$ = 60 mA to compensate for the propagation loss.

Below threshold, amplified spontaneous emission circulating in the ring cavity is periodically enhanced at the longitudinal resonances, producing pronounced modulation of the optical spectrum. The modulation depth increases as the round-trip net gain approaches the lasing threshold and can therefore be used to extract the net modal gain using the Hakki–Paoli method[1]. For each longitudinal mode, the spectral modulation ratio is defined as

$$S(\lambda) = \frac{I_{\max}(\lambda)}{I_{\min}(\lambda)} \tag{1}$$

where $I_{\max}$ and $I_{\min}$ are the resonance-peak intensity and the intensity at an adjacent minimum. Since the power coupling efficiency of the dual-ring structure is only 1% and the loss ring does not lase, the waveguide loss can be approximately estimated using a single-ring model. For a single-ring response, the net modal gain is:

$$g_{net} = \frac{2}{L}\ln\left(\frac{\sqrt{S}-1}{\sqrt{S}+1}\right) - \frac{1}{L}\ln T_{rt} \tag{2}$$

Where $L$ is the ring circumference, $T_{\mathrm{rt}}$ is the round-trip power transmission associated with discrete losses, and for the power coupling efficiency of 0.1 between the gain ring and SOA waveguide, $T_{\mathrm{rt}}$ = 0.9. The extracted net modal gain spectra are shown in Supplementary Fig. S1b. The effective internal modal loss used in the analysis is $\alpha_{\mathrm{i}} = 3 \pm 0.5\ \mathrm{cm}^{-1}$.

The approximately linear gain–current dependence below threshold was used to map $I_L$ onto the coefficients of the coupled-mode model. In this calibration, the gain-ring coefficient was held approximately clamped while the loss-ring coefficient varied with $I_{\mathrm{L}} < I_{\mathrm{th}}$.

$$\Delta g_{G,L} \simeq \alpha_i\left(1 - \frac{I_L}{I_{th}}\right) \tag{3}$$

The resulting eigenvalues are shown in Supplementary Fig. S1c, d. At zero relative detuning, increasing $I_L$ reduces the gain–loss contrast. The imaginary eigenvalue branches merge at the exceptional point (EP), while the real-frequency branches split on the higher-current side. The estimated EP occurs at $I_L \approx 67$ mA. The measured locking interval of 72–90 mA lies on the higher-current, unbroken side of this zero-detuning model.

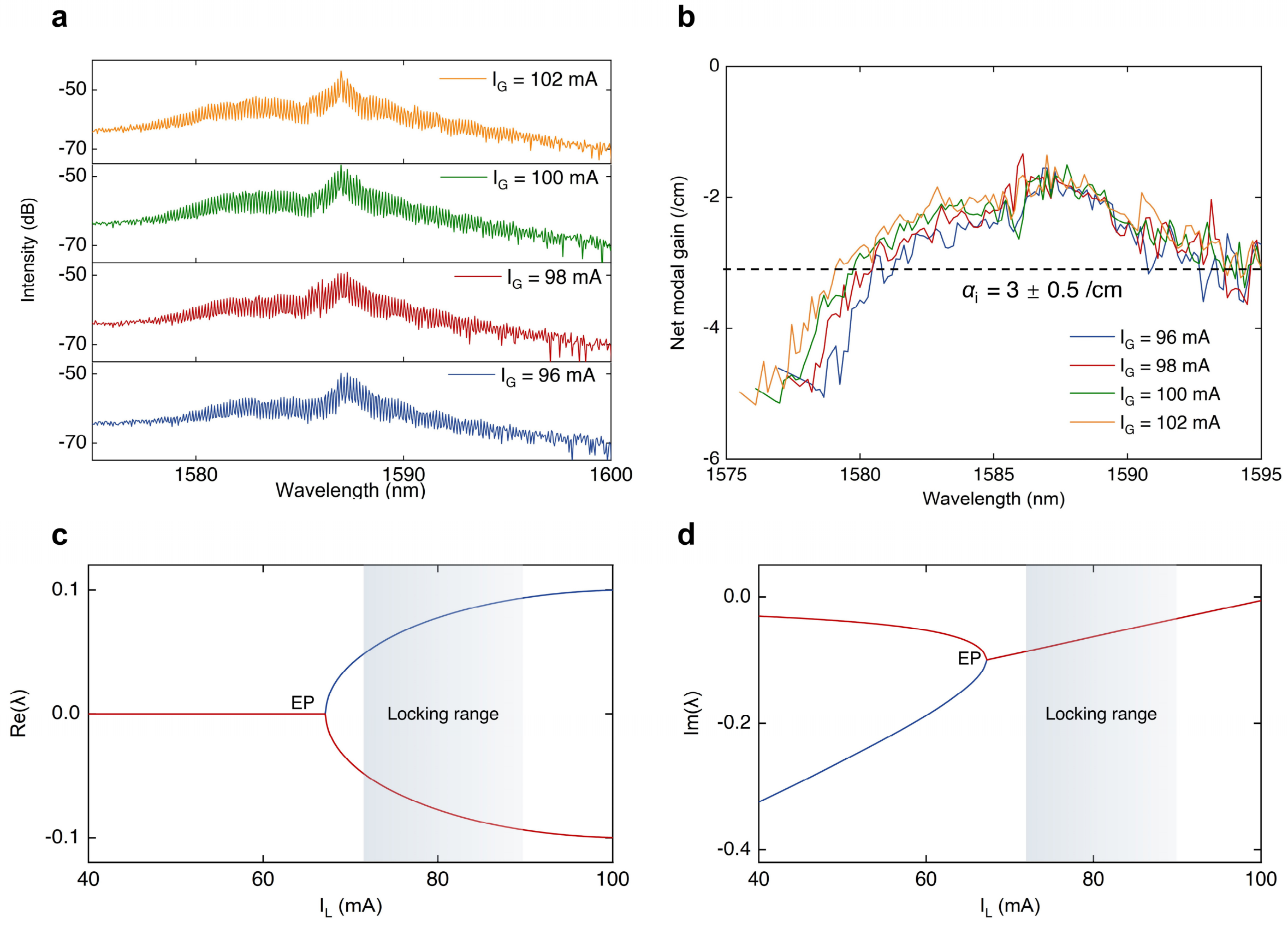

**Fig. S1: Experimental characterization of the modal gain and non-Hermitian eigenvalues. a**, Optical spectra at $I_G$ = 96, 98, 100 and 102 mA, with $I_L$ = 75 mA and $I_{SOA}$ = 60 mA. The nominal lasing threshold is $I_{th}$ = 100 mA. **b**, Net modal gain obtained from the spectral modulation using the ring-adapted Hakki–Paoli relation. The dashed line marks $-\alpha_i$ for the effective internal modal loss $\alpha_i = 3 \pm 0.5$ cm$^{-1}$. **c, d**, Calculated real (c) and imaginary (d) parts of the coupled-ring eigenvalues as functions of $I_L$, using the gain–current calibration described in Eq. (3). Blue and red curves show the two eigenvalue branches. The EP is estimated at $I_L \approx 67$ mA. Shading marks the measured locking interval of 72–90 mA.

## Supplementary Note 2: Theoretical model of non-Hermitian mode locking

### *Nonlinear non-Hermitian supermode restructuring*

For identical rings at zero relative detuning and equal net gain, the coupling-induced supermode frequency splitting is

$$\Delta f_{SM} = \frac{\arcsin(\kappa_c)}{\pi}\mathrm{FSR} \tag{4}$$

Here, $\kappa_c$ is the field coupling coefficient and $K = \kappa_c^2$ is the corresponding power coupling coefficient. For $K = 0.01$, $\kappa_c = 0.1$. With a free spectral range (FSR) of 18.11 GHz, Eq. (4) gives $\Delta f_{SM} \approx 577$ MHz, compared with the measured splitting of approximately 542 MHz. For one pair of longitudinal modes, the effective non-Hermitian Hamiltonian is:

$$H = (\bar{\omega} + i\bar{\sigma})I + \begin{pmatrix} \Delta\omega_0/2 + i\gamma & \kappa \\ \kappa & -\Delta\omega_0/2 - i\gamma \end{pmatrix} \tag{5}$$

Here, $\bar{\omega}$ is the mean resonance angular frequency, $\bar{\sigma} = (\sigma_G + \sigma_L)/2$ is the mean net field growth rate, and $\gamma = (\sigma_G - \sigma_L)/2$ is half the growth-rate difference between the rings. The coupling rate κ has units of angular frequency, with $\kappa = \pi\Delta f_{SM} = \mathrm{FSR}\arcsin(\kappa_c)$ in the symmetric limit. We use the convention $A(t) \propto \exp(-i\lambda t)$, so positive $\mathrm{Im}(\lambda)$ denotes field growth. The corresponding complex eigenfrequencies are

$$\lambda_\pm = \bar{\omega} + i\bar{\sigma} \pm \sqrt{\kappa^2 + \left(\frac{\Delta\omega}{2} + i\gamma\right)^2} \tag{6}$$

The total relative detuning is $\Delta\omega = \Delta\omega_0 + \Delta\omega_{NL}$, where $\Delta\omega_0$ is the initial resonance mismatch between the two rings, and $\Delta\omega_{NL}$ is the nonlinear frequency detuning. Mode locking therefore requires spectral alignment of the longitudinal modes in the gain and loss rings, which can be achieved in MQW lasers through current-induced refractive-index tuning. For a stationary lasing supermode, gain saturation imposes

$$\mathrm{Im}(\lambda_{\mathrm{lasing}}) \approx 0 \tag{7}$$

This condition applies to the complete coupled mode and does not require the net gain of each ring to vanish separately. When both rings' current is larger than the threshold ($\gamma \to 0$), real detuning produces no imaginary eigenvalue splitting in Eq. (6). Finite gain–loss asymmetry is therefore required for the detuning-induced modal discrimination considered here.

At static resonance matching, $\Delta\omega_0 = 0$ and $\Delta\omega = \Delta\omega_{NL}$, the gain ring operates in the CW lasing regime before the onset of mode locking, while the loss ring remains below threshold. The resulting intracavity peak-power contrast or carrier fluctuation is small, such that $\Delta\omega_{NL} \ll \gamma$. Equation (6) can be simplified to

$$\lambda_\pm \approx \bar{\omega} + i\bar{\sigma} \pm \sqrt{\kappa^2 - \gamma^2} \pm \frac{i\gamma\Delta\omega_{NL}}{2\sqrt{\kappa^2 - \gamma^2}} \tag{8}$$

(9)

We define the nonlinear Hamiltonian modal at unbroken discrimination as

$$D_{NH} = \left|\mathrm{Im}(\lambda_+) - \mathrm{Im}(\lambda_-)\right| \approx \frac{\left|\gamma \Delta \omega_{NL}\right|}{\sqrt{\kappa^2 - \gamma^2}} \approx S_{NH}\left|\Delta \omega_{NL}\right| \tag{10}$$

Here, "unbroken" refers to the zero-detuning Hamiltonian after subtraction of the common frequency $\bar{\omega}$ and growth-rate term $\bar{\sigma}$. $S_{NH} = |\gamma|/\sqrt{(\kappa^2 - \gamma^2)}$ is the small-detuning sensitivity. Figure S2a shows the nonlinear modal discrimination coefficient, $S_{NH}$, as a function of $I_L$. The sensitivity increases towards the EP from the unbroken side. This enhancement identifies a regime of strong nonlinear modal discrimination; it quantifies the sensitivity of the imaginary eigenvalue splitting to weak nonlinear detuning under continuous-wave conditions, although it does not by itself establish phase locking between longitudinal modes. For $|\Delta\omega| \gg \kappa$ and $|\gamma|$, the eigenmodes approach the individual-ring modes and inter-ring hybridization weakens. Pulse formation is therefore assessed with the time-domain model below.

***Kerr- and carrier-induced nonlinear detuning***

The differential Kerr detuning between the two rings is:

$$\Delta\omega_K = -\mathrm{FSR}\,\gamma_{\mathrm{NL}} L (P_G - P_L) \tag{11}$$

Here, $P_G$ and $P_L$ are the instantaneous intracavity powers of the gain-ring and loss-ring, respectively; $\gamma_{NL} = 2\pi n_2/(\lambda A_{eff})$ is the nonlinear propagation coefficient, and $L$ is the ring circumference. The quantities $n_2$ and $A_{eff}$ denote the nonlinear refractive index and effective mode area, respectively. With FSR = $1/T_R$, a round-trip phase shift produces an angular-frequency shift equal to minus that phase shift divided by $T_R$. Carrier-induced refractive-index changes provide an additional contribution through the linewidth-enhancement factor $\alpha_H$:

$$\Delta\omega_H = \frac{\mathrm{FSR}\,\alpha_{\mathrm{H}} L}{2}\delta g_c \tag{12}$$

To describe this response, we use a spatially uniform carrier-density model with a single recovery time. For intracavity power $P$(t), the rate equation is[2].

$$\frac{dN}{dt} = \frac{I}{qV} - \frac{N}{\tau_c} - \frac{gP(t)}{\Gamma A_{sat}\hbar\omega_0} \tag{13}$$

Here, $q$ is the elementary charge, $I$ is the injection current, $V$ is the active volume, and $A_{sat} = V/\Gamma L$ is the effective active cross-sectional area. The modal gain is $g = \Gamma\alpha_0(N - N_0)$, with confinement factor $\Gamma$, material differential gain $\alpha_0$, and transparency density $N_0$. Using a carrier lifetime $\tau_c$ = 0.2–0.3 ns gives the equivalent gain equation

$$\frac{dg}{dt} = \frac{g_p - g}{\tau_c} - \frac{g}{E_{\mathrm{sat}}}P(t) \tag{14}$$

Here, $g_p$ is the unsaturated modal gain at the applied current. Under this normalization, $E_{sat}$ = $=h\nu A_{sat}/\alpha_0$. For $A_{sat} \approx 1$ µm$^2$ and $\alpha_0 = 3 \times 10^{-20}$ m$^2$, $E_{sat} \approx 4$ pJ at $\lambda = 1600$ nm.

The response to small fluctuations around a continuous-wave (CW) state differs from carrier depletion by an established pulse. Linearizing Eq. (14) about the steady-state gain $\bar{g}$ and power $\bar{P}$ gives:

$$\frac{d\delta g}{dt} + \left(\frac{1}{\tau_c} + \frac{\bar{P}}{E_{\mathrm{sat}}}\right)\delta g = -\frac{\bar{g}}{E_{\mathrm{sat}}}\delta P \tag{15}$$

The corresponding effective carrier response time is:

$$\tau_{\text{eff}} = \left( \frac{1}{\tau_c} + \frac{\bar{P}}{E_{\text{sat}}} \right)^{-1} \tag{16}$$

For a CW intracavity power of 3 – 4 mW, the stimulated-depletion contribution to the relaxation rate is small, giving $\tau_{\text{eff}} \approx \tau_{\text{c}}$. It follows that, under CW conditions, the response time to small fluctuations in intensity is limited by the carrier lifetime. Transitioning from CW mode to pulsed mode-locking requires rapid discrimination between high-intensity optical pulses and low-intensity noise. Carrier-induced nonlinear detuning makes it difficult to trigger the transition into a fast pulse. In contrast, the Kerr-induced detuning follows the intracavity power variation essentially instantaneously. Certainly, this timescale comparison alone does not exclude a carrier contribution to pulse formation. During an established pulse much shorter than $\tau_{\text{c}}$, injection and recovery can be neglected, giving:

$$g(t) \simeq g_b \exp\left[ -\frac{E(t)}{E_{sat}} \right] \quad E(t) = \int_{t_0}^{t} P(t')dt' \tag{17}$$

Here, $g_{\text{b}}$ is the gain immediately before the pulse, and E($t$) is the energy accumulated from the pulse leading edge to time $t$. After the pulse, $E(t) = E_{\text{p}}$ and $g_{\text{after}} = g_{\text{b}} \exp(-E_{\text{p}}/E_{sat})$ is the gain after the pulse. The gain perturbation is $\delta g(\text{t}) = g(t) - g_{\text{b}}$. Carrier-induced chirp can therefore develop during the pulse, while recovery follows the slower carrier dynamics. This response differs from instantaneous intensity-dependent transmission.

***Time-domain model of the coupled semiconductor rings***

We model the slowly varying fields $A_{\text{G}}$ and $A_{\text{L}}$ using coupled travelling-wave equations, with $|A_{\text{j}}|^2 = P_{\text{j}}$. For j $\in \{G, L\}$.

$$\frac{\partial A_j}{\partial z} = \left[ \frac{1}{2}\left(g_j - \alpha_j\right) - i\frac{\alpha_{H,j}}{2}\delta g_{c,j} \right] A_j - i\frac{\beta_{2,j}}{2}\frac{\partial^2 A_j}{\partial \tau^2} + i\gamma_{\text{NL},j} \, | A_j |^2 \, A_j + i\kappa_z A_k \tag{18}$$

Here, $z$ is the propagation coordinate, and $\tau$ is retarded time. The coefficients $g_j$, $\alpha_j$ denote modal gain and internal loss, respectively. The carrier phase term uses the modal-gain perturbation $\delta g_{\text{c,j}}$; $\beta_{2,\text{j}}$ is the group-velocity dispersion, $\alpha_{\text{H}}$ is the linewidth-enhancement factor, and $\kappa_{\text{z}}$ is the coupling rate per unit length. Finite gain bandwidth is included through spectral filtering in the numerical model. The relative contribution of dispersion can be estimated from

$$R_D = \frac{|\beta_2| L}{\tau^2} \tag{19}$$

For an intensity profile proportional to $\text{sech}^2(t/\tau_0)$, the 7.9 ps intensity full width at half maximum (FWHM) corresponds to $\tau_0 \approx 4.48$ ps. The reported estimate $R_{\text{D}} \approx 3.8 \times 10^{-4}$ indicates weak dispersion over one round trip. A small $R_{\text{D}}$ does not alone establish that dispersion is negligible over the full pulse build-up time.

We first isolate Kerr-induced nonlinear phase modulation in the time-domain model. The steady-state pulse is concentrated in the gain ring, with a weaker pulse in the loss ring (Fig. S2b). Starting from a gain-ring average power of 40 mW, the peak powers evolve towards distinct steady values (Fig. S2c). The temporal evolution shows pulse formation over several hundred round trips, followed by persistence throughout the remaining simulation (Fig. S2d). These results demonstrate pulse formation within the chosen model and parameter set.

We then include carrier-induced phase modulation to examine its effect on the calculated spectrum and pulse profile. The carrier-inclusive calculation uses $E_p/E_{sat} = 0.5$. Supplementary Fig. S2e, f compares the Kerr-only and Kerr-plus-carrier results. The spectral FWHM increases from 6.71 to 7.07 nm, while the FWHM increases from 6.5 to 7.8 ps. These changes describe pulse reshaping after carrier-induced phase modulation is included.

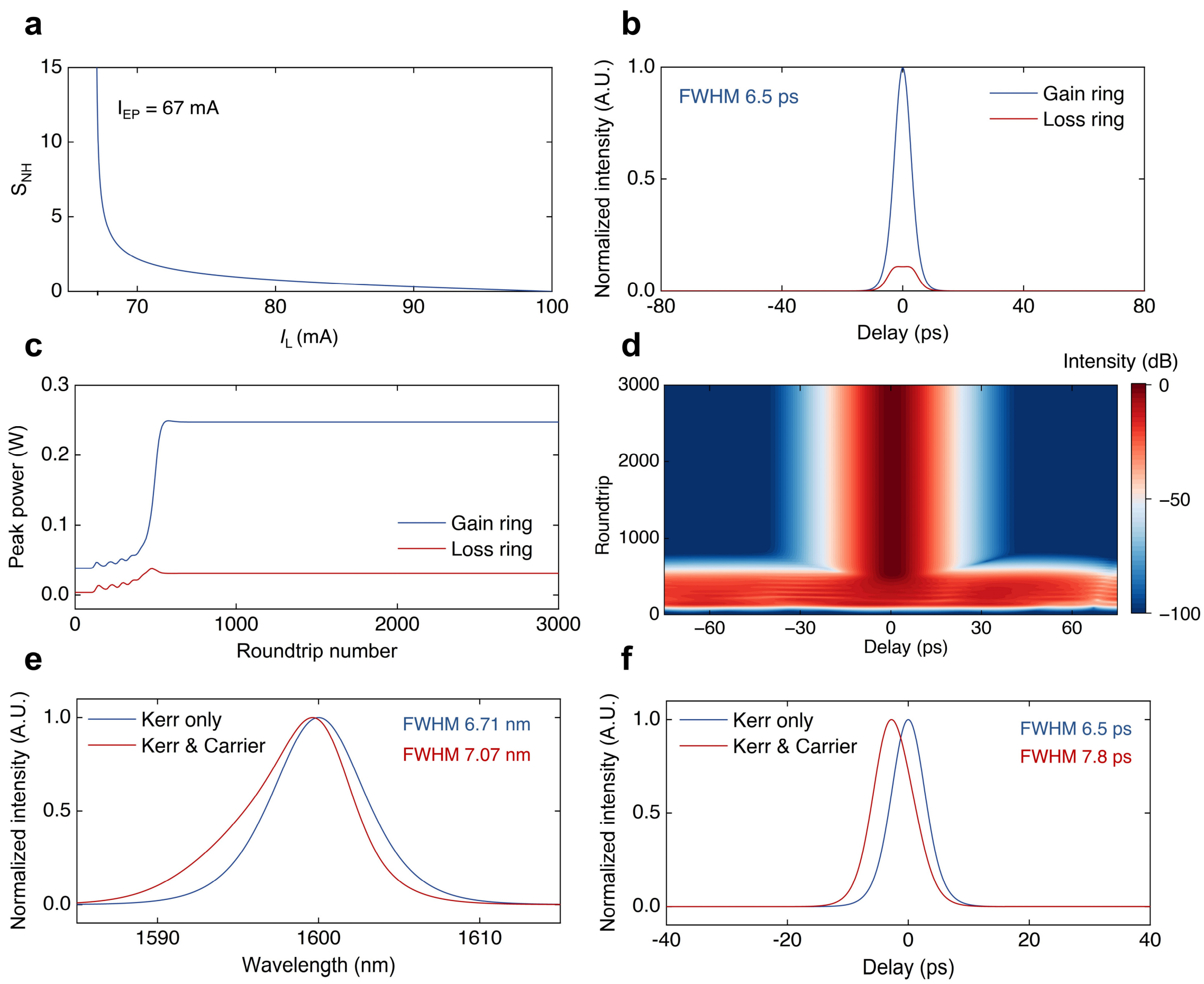

**Fig. S2: Nonlinear modal discrimination and simulated pulse dynamics.** **a**, Small-detuning sensitivity $S_{NH}$ as a function of $I_L$ on the unbroken side of the estimated EP at 67 mA. The first-order approximation is not valid at the EP. **b–d**, Simulations with Kerr-induced nonlinear phase modulation. **b**, Steady-state intensity profiles in the gain ring (blue) and loss ring (red); the gain-ring pulse has an intensity FWHM of 6.5 ps. **c**, Evolution of the peak powers in the two rings. **d**, Intracavity intensity versus delay and round-trip number, with start gain-ring average power of 40 mW, showing pulse build-up and subsequent persistence to 3000 round trips. **e, f**, Calculated spectra (**e**) and pulse profiles (**f**) with Kerr modulation alone (blue) and with Kerr and carrier-induced modulation (red). The spectral FWHM values are 6.71 and 7.07 nm, and the corresponding pulse intensity FWHM values are 6.5 and 7.8 ps.

**TABLE S1: Simulation parameters.**

| Parameter | Symbol | Value | Unit |
|---|---|---|---|
| Power coupling coefficient | $K$ | 0.01 | |
| Field coupling coefficient | $\kappa_c$ | 0.1 | |
| Free spectral range | FSR | 18.1 | GHz |
| Kerr coefficient | $n_2$ | $1.5\times10^{-17}$ | $m^2/W$ |
| Effective mode area | $A_{eff}$ | 3.8 | $\mu m^2$ |
| Linewidth-enhancement factor | $\alpha_H$ | 3 | |
| Inverse of the group velocity | $\beta_1$ | $1.17\times10^{-8}$ | s/m |
| GVD | $\beta_2$ | -1.6 | $ps^2/m$ |
| Carrier lifetime | $\tau_N$ | 0.25 | ns |
| Ring length | $L$ | 4.718 | mm |
| Static detuning | $\Delta\omega_0$ | 0 | /s |
| Simulation time window | $T_W$ | 200×FSR | ps |
| Gain saturation energy | $E_{sat}$ | 4 | pJ |
| Average power | $\overline{P}$ | 40 | mW |
| Material differential gain | $\alpha_0$, | $3 \times 10^{-20}$ | $m^2$ |
| Transparency density | $N_0$ | $1 \times 10^{24}$ | $m^{-3}$ |
| Confinement factor | $\Gamma$ | 0.05 | |

## Supplementary Note 3: Electrical tuning of the mode-locking window

We swept $I_L$ from 0 to 150 mA at fixed $I_G$ values from 410 to 500 mA in 10 mA steps. The resulting optical spectra are shown in Supplementary Fig. 3. Finite locking intervals were identified at each gain-ring current, with boundaries that varied non-monotonically with $I_G$. At $I_G$ = 490 and 500 mA, two separate intervals were observed. Electrical injections change both the gain and resonance frequencies, so successive longitudinal-mode pairs may approach resonance as the currents vary. This provides a possible explanation for the separated windows, although the spectral maps do not directly measure the inter-ring detuning. Across the maps, the marked intervals span $I_L$ = 68.25–97.5 mA. They lie above the estimated 67 mA EP of the zero-detuning calibration. This comparison is consistent with operation on the unbroken side of the reference model, subject to its gain-clamping and detuning assumptions.

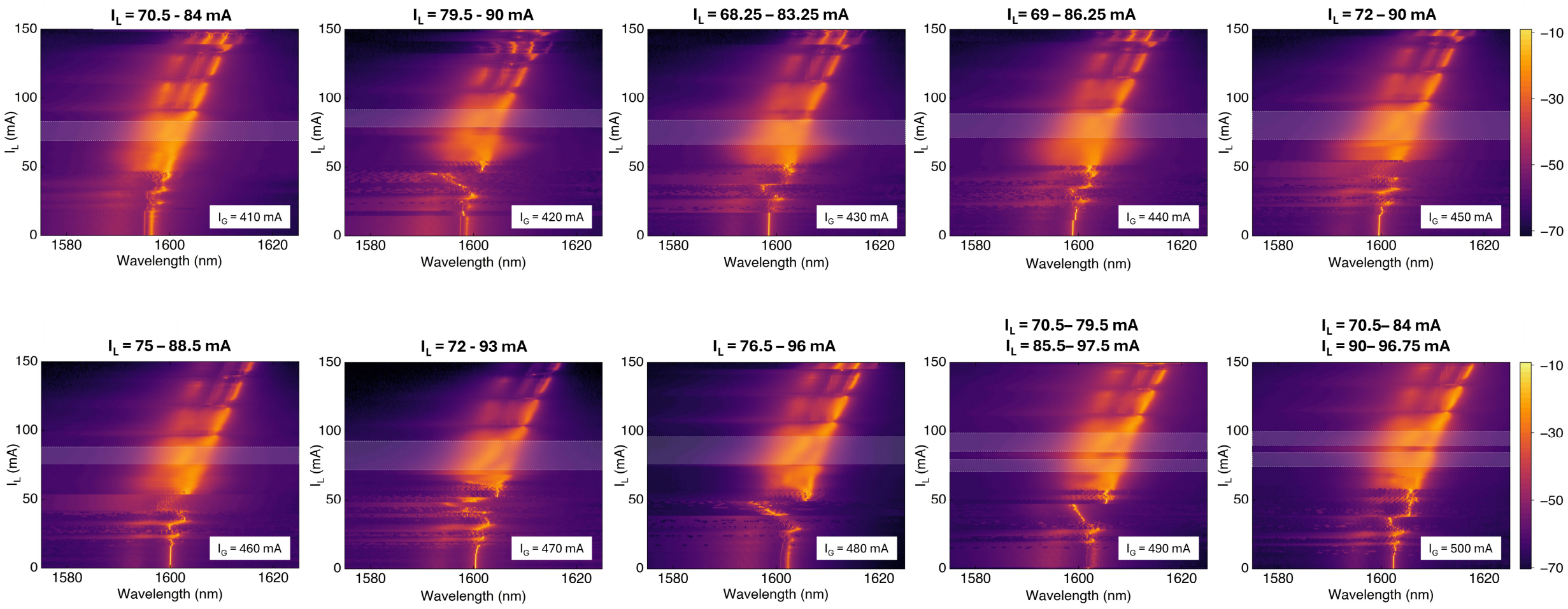

**Fig. S3: Evolution of the mode-locking window with gain-ring current.** Electrical tuning of the locking intervals. Optical spectra recorded while sweeping $I_L$ from 0 to 150 mA at fixed gain-ring currents. From left to right, $I_G$ = 410–450 mA in the top row and 460–500 mA in the bottom row, in 10 mA steps. Shaded horizontal bands mark the identified locking intervals; their current bounds are indicated above each panel. Two separated intervals occur at $I_G$ = 490 and 500 mA. All maps use the same colour scale.

## Supplementary Note 4: Fabrication process of non-Hermitian dual-ring laser

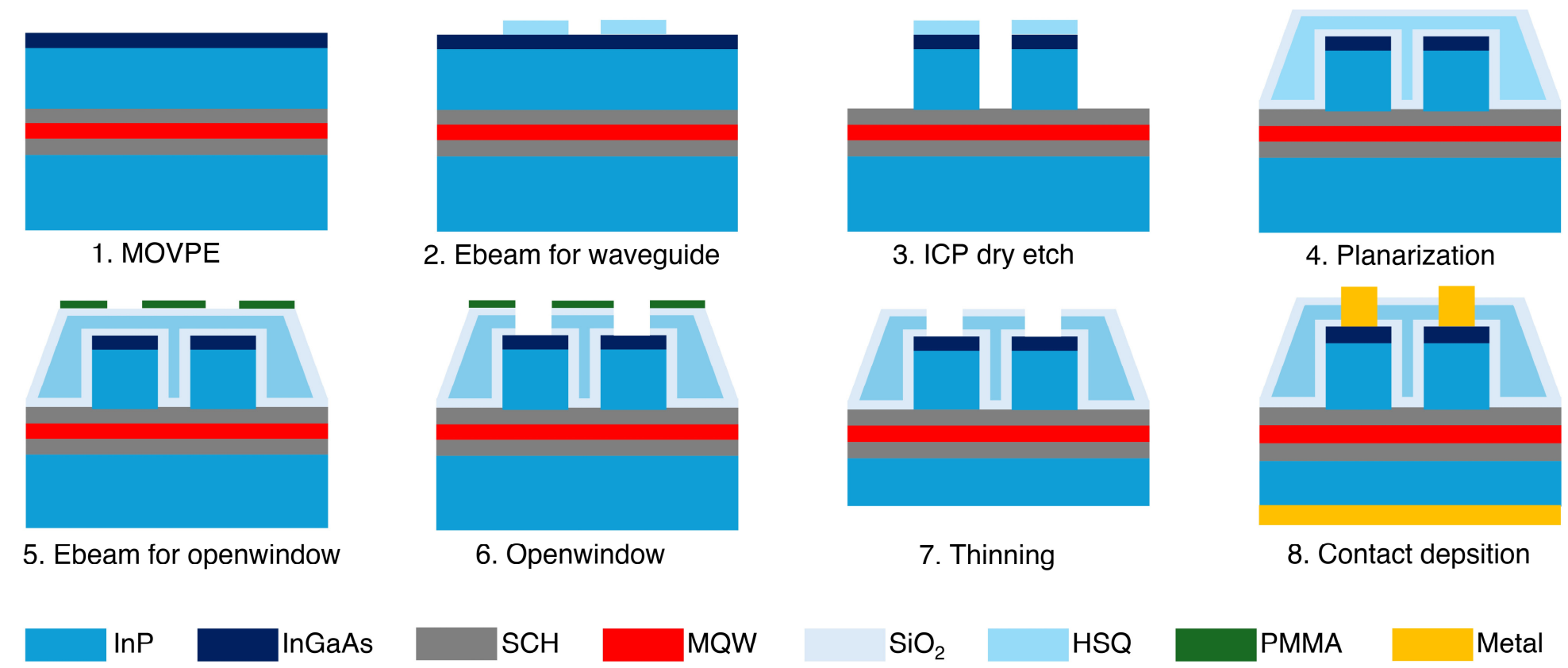

**Fig. S4.** Fabrication process of dual-ring laser. Schematic cross-sections show epitaxial growth, electron-beam patterning of the waveguides, inductively coupled plasma (ICP) dry etching and planarization. Subsequent steps comprise electron-beam patterning and opening of contact windows, substrate thinning, and metal contact deposition. Colours identify InP, InGaAs, separate-confinement heterostructure (SCH) layers, multiple quantum wells (MQWs), $SiO_2$, hydrogen silsesquioxane (HSQ), poly (methyl methacrylate) (PMMA) and metal. MOVPE denotes metal–organic vapour-phase epitaxy.